\documentclass[showpacs,twocolumn,amsmath,amssymb,fixfloats]{revtex4}

\usepackage[english]{babel}
\usepackage{amsmath}
\usepackage{amssymb}
\usepackage{graphicx}

\newcommand{\ket}[1]{|#1\rangle}
\newcommand{\bra}[1]{\langle#1|}

\begin{document}

\title{Cavity QED determination of atomic number statistics in optical lattices}

\author{W. Chen}
\author{D. Meiser}
\author{P. Meystre}

\affiliation{Department of Physics, The University of Arizona,
1118 E. 4th Street, Tucson, AZ 85721 }

\begin{abstract}
We study the reflection of two counter-propagating modes of the
light field in a ring resonator by ultracold atoms either in the
Mott insulator state or in the superfluid state of an optical
lattice. We obtain exact numerical results for a simple two-well
model and carry out statistical calculations appropriate for the
full lattice case. We find that the dynamics of the reflected
light strongly depends on both the lattice spacing and the state
of the matter-wave field. Depending on the lattice spacing, the
light field is sensitive to various density-density correlation
functions of the atoms. The light field and the atoms become
strongly entangled if the latter are in a superfluid state, in
which case the photon statistics typically exhibit complicated
multimodal structures.
\end{abstract}
\maketitle

\section{\label{introduction} Introduction}

The study of ultracold atoms in optical lattices has been a very
active field of research both experimentally and theoretically in
recent years. One motivation to study these systems is that they
provide clean realizations of important models of condensed matter
physics \cite{lewensteincondmat} such as the Bose-Hubbard and
Fermi-Hubbard models \cite{Jaksch:BECinLattice,Zoller:OLreview},
spin systems \cite{GarciaRipoll:SpinHamiltonians} and the Anderson
lattice model \cite{Miyakawa:AndersonModel}. A particularly
well-known example is the prediction \cite{Jaksch:BECinLattice} and observation
of the Mott-insulator to superfluid transition
\cite{Bloch:MottInsulator1,Bloch:MottInsulator2} in trapped
Rubidium atoms. An important advantage of ultracold atoms over
conventional solid-state systems is that they offer an exquisite
degree of control over system parameters such as interaction
strengths, densities and tunneling rates. Ultracold atomic systems
are also very versatile in that the atoms involved can be either
fermionic or bosonic, and can be associated into molecules of
either statistics
\cite{Rom:MoleculesLattices,Stoeferle:MoleculesLattice,Thalhammer:MoleculesLattice,Jaksch:creationmoleculeinOL}.
Furthermore the dimensionality of the systems can be tuned from
three to two and one-dimensional. Important other potential
applications of ultracold atoms in lattices include quantum
information
\cite{Jaksch:entanglementlattices,Mandel:entanglementlattices} and
improved atomic clocks \cite{Takamoto:LatticeClock}.

The ability to characterize the many-particle state of atomic
fields is an important ingredient of many of these studies. While
a full characterization of the field would ideally be desirable, a
great deal can already be learned from the atomic density
fluctuations at each lattice site and from the intersite density
correlations. The counting statistics of atoms in an optical
lattice have previously been studied by using spin changing
collisions \cite{Gerbier:NumberSqueezingLattice}, atomic
interferences in free expansion \cite{Roberts:ProbingMott} and the
conversion of atoms into molecules via photoassociation
\cite{Ryu:photoassociationlattice} or Feshbach resonances
\cite{Stoeferle:MoleculesLattice}. The main goal of this paper is
to propose and analyze an alternative cavity-QED based method to
measure these properties.

Our proposed scheme is an extension to trapped ultracold atoms of
the familiar diffraction technique used to probe order in
crystalline structures. One important new aspect is that since we
wish to measure the quantum fluctuations of the atomic density we
cannot simply use classical radiation, which only provides
information on some average of the atomic occupation numbers.
Instead, we exploit the quantum nature of the light field and
the fact that it can become entangled with the
atoms to probe the number statistics of the matter-wave
field. It is for this reason that a cavity-QED geometry is
particularly attractive: as is well known, high-$Q$ optical
cavities can significantly isolate the system from its
environment, thus strongly reducing decoherence and ensuring that
the light field remains quantum mechanical for the duration of the
experiment.

More specifically, we consider two counter-propagating field modes
in a high-$Q$ ring cavity, coupled via Bragg scattering off the
atoms trapped in an optical lattice. If one of the
counter-propagating modes is initially in vacuum, this is
analogous to a quantum version of the Bragg reflection of X-rays
by a crystal.

The remainder of this article is organized as follows: Section
\ref{model} describes our model and presents several general
results of importance for the following analysis. In particular,
it shows that the dynamics of the light field strongly depends on
the manybody state of the atomic field as well as on the lattice
spacing. Section \ref{twowellcase} presents a series of results
for the case of a simple two-well system, analyzing the properties
of the Bragg-reflected light field for atoms in a Mott insulator
state and for a superfluid described both in terms of a
number-conserving state and of a mean-field coherent state. In
particular, we find that these two descriptions lead to major
differences in the properties of the scattered light. Section
\ref{latticecase} then turns to the case of a large lattice.
Finally, section \ref{conclusion} is a conclusion and outlook.

\section{\label{model} Model}

\begin{figure}
\includegraphics{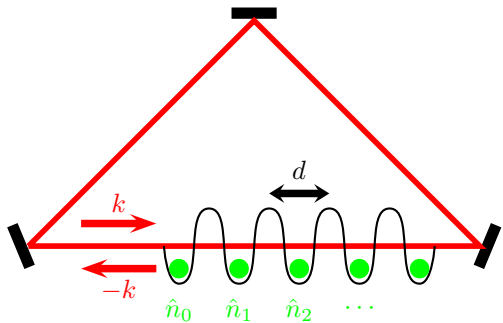}
\caption{(Color online) Atoms trapped in an optical lattice with lattice constant
$d$ in a ring resonator and interacting with two
counter-propagating modes with wave vectors $\pm k$.}
\label{schematic}
\end{figure}

We consider a sample of bosonic two-level atoms with transition
frequency $\omega_a$ trapped in the lowest Bloch band of a
one-dimensional optical lattice with $M$ lattice sites and a
lattice spacing $d$, see Fig. \ref{schematic}. We assume for
simplicity that the effects of tunneling and collisions are fully
accounted for by the initial state of the matter-wave field, and
neglect them during the subsequent scattering of the weak
quantized probe fields, taken to be two counter-propagating
(plane-wave) modes of a ring cavity with mode functions $\mathcal
E_{\pm k}(z)$, wave-vectors $\pm k$ and frequency $\omega_k$. We
further assume that these fields are far detuned from the atomic
transition, $|\Delta|\equiv|\omega_k-\omega_a|\gg \gamma,
\Omega_R$, with $\gamma$ the atomic linewidth and $\Omega_R$ the
vacuum Rabi frequency, so that the excited electronic state of the
two-level atoms can be adiabatically eliminated.

Expanding the field operators for the atomic excited state $\hat
\psi^{(e)}(z)$ and ground state $\hat \psi^{(g)}(z)$ in the
Wannier basis of the lowest Bloch band as
$$
\hat \psi^{(e,g)}(z)=\sum_{m=0}^{M-1} \psi_m^{(e,g)}(z)\hat c_m^{(e,g)},
$$
where $\hat c_m^{(e)}$ and $\hat c_m^{(g)}$ are the bosonic
annihilation operators for an atom in the excited and ground state
at lattice site $m$ and $\psi_m^{(e)}$ and $\psi_m^{(g)}$
the corresponding wave functions, this system is described by the
effective Hamiltonian
\begin{eqnarray}
\hat H&=& \hbar g[ \hat N(0)(\hat a_k^\dagger \hat a_k + \nonumber
\hat a_{-k}^\dagger \hat a_{-k})\nonumber\\&&
 +\hat N(d)
    \hat a_{-k}^\dagger\hat a_k +\hat N(-d)\hat a_k^\dagger \hat a_{-k}],
    \label{effectivehamiltonian}
\end{eqnarray}
where we have neglected collisions and tunneling, as already
discussed, and also ignore all loss processes. In this expression
\begin{equation}
\label{Nofd} \hat N(d)=\sum_{m=0}^{M-1} e^{2imkd}\hat
c_m^{(g)\dagger} \hat c_m^{(g)}
\end{equation}
and the coupling frequency is
\begin{equation}
g=\frac{2\wp^2}{\Delta \hbar^2}\left|\int dz
\mathcal E_{\pm k}(z)\psi_{0}^{(e)*}(z)\psi_0^{(g)}(z)\right|^2,
\end{equation}
with $\wp$ the dipole matrix element of the atomic transition.
Since only the ground electronic state is relevant in this model,
we drop the superscript $(g)$ in the following for notational
clarity.

Our proposed detection scheme relies crucially on the observation
of the coherent dynamics resulting from the Hamiltonian
(\ref{effectivehamiltonian}), hence it is important that the
associated characteristic time scales are significantly faster
than than those associated with losses and with the atomic motion
in the lattice. To illustrate that this is within current
experimental reach, we consider explicitly the experimental
parameters of Klinner {\it et. al.}
\cite{Klinner:NormalModeSplitting}. Assuming that the atoms are
confined to much less than an optical wavelength at the antinodes
of the lattice potential, we can approximate the coupling constant
between the cavity modes and a single atom as
$$
g\approx \frac{\wp^2\omega_k}{\Delta\hbar\epsilon_0V},
$$
where $V$ is the mode volume of the cavity and $\wp\approx 2.3\times 10^{-29}$
Cm is the relevant dipole matrix element for the $D_2$ line of ${}^{85}$Rb. For
a detuning of $\Delta=-1$ GHz from that transition, a cavity length of $0.1$ m,
and a mode waist of $100\;\mu$m we find $g\approx 50$ s${}^{-1}$.

As we show later on, the fluctuations of the operator $\hat N(d)$
are central in the determination of the spectral properties of
$\hat G$. As these typically scale like the square root of the
total number of atoms we find that for a sample of $10^6$ atoms
the relevant frequencies for the coherent evolution are of the
order of 50 kHz. Such frequencies are several times larger than
the decay rate of state-of-the-art optical cavities with large
mode volume, 17 kHz in the experiments of Klinneris  {\it et. al.}
The spontaneous emission from the atomic excited state is
negligible at this detuning. Indeed, the fact that in the
experiment of Klinner a splitting of the normal modes has been
observed is direct experimental evidence that the coherent
dynamics can be made dominant over the losses.

Turning now to the characteristic time scales for the
center-of-mass atomic dynamics, specifically the interwell
tunneling rate and the two-body collision rate, they can be
controlled with high accuracy, respectively through the depth of
the lattice potential and via a magnetic Feshbach resonance. It is
therefore possible to operate under experimental conditions such
that both are much smaller than the coupling strength between
atoms and light. This justifies ``freezing'' the atomic dynamics
resulting from tunneling and collisions once the probe fields are
turned on.

It is well known that, depending upon the ratio between tunneling
and two-body collisions, bosonic atoms in an optical lattice can
undergo a transition from a superfluid to a Mott insulator state.
With this in mind, we consider initial atomic states that
correspond to these two extreme situations, specifically: (a) a
Mott insulator state with a well-defined atom number in each well,
(b) a state where each well is in a coherent state, a mean-field
approximation of the superfluid state, and (c) a more realistic
description of atoms in the superfluid state with a fixed total
number of atoms $N$. These three states are given explicitly by
\begin{equation}
\ket{\psi_{\rm Mott}}=\ket{n_0,n_1,\ldots,n_{M-1}},
\end{equation}
\begin{equation}
\ket{\psi_{\rm SF1}}=\ket{\alpha_0,\alpha_1,\ldots,\alpha_{M-1}},
\end{equation}
where $|\alpha_m|^2$ is the mean number of atoms in well $m$, and
\begin{equation}
\ket{\psi_{\rm SF2}}=\mathcal N^{-1}\left(\sum_{m=0}^{M-1}\ {\hat
c}_m^\dagger\right)^N\ket{0},
\end{equation}
where $\mathcal N=\sqrt{N!M^N} $ is a normalization constant.

We further assume for simplicity that the mode propagating in the
$-k$ direction along the ring is initially in a vacuum while the
other mode is in a Fock state with $n_{\rm tot}$ photons. Our
results are qualitatively independent of the exact state of the
light field, but we will point out the modifications brought about
by a coherent state instead of a number state when appropriate.

The intensities of the two counter-propagating modes, $\langle \hat
a_{k}^\dagger\hat a_{k}\rangle$ and $\langle \hat a_{-k}^\dagger\hat
a_{-k}\rangle$, are easily obtained from the solution of the Heisenberg
equations of motion for ${\hat a}_{\pm k}(t)$,
\begin{widetext}
\begin{equation}
\left[
\begin{array}{c}
\hat a_{-k}(t)\\
\hat a_{k}(t)
\end{array}
\right]
=
e^{-ig\hat N(0)t}
\left[
\begin{array}{cc}
\cos \hat G(d)t &
-i\hat Q(d) \sin \hat G(d)t\\
-i\hat Q^\dagger(d) \sin \hat G(d)t&
\cos \hat G(d)t
\end{array}
\right]
\left[
\begin{array}{c}
\hat a_{-k}(0)\\
\hat a_{k}(0)
\end{array}
\right],
    \label{formalsolution}
\end{equation}
\end{widetext}
where
\begin{equation}
\hat Q(d)=\frac{\hat N (d)}{\sqrt{\hat N(d)\hat N(-d)}}
\end{equation}
is unitary and we have introduced the operator
\begin{equation}
\hat G(d)=g\sqrt{\hat N(d)\hat N(-d)}
\end{equation}
describing the coupling between the two modes.
The closed form solution (\ref{formalsolution}) follows from the
observation that in the absence of tunneling the optical field
only couples to conserved quantities of the atomic field. Note
however that the number statistics of the atoms at each site is
conserved for many cases of practical interest even in the
presence of tunneling. This is for instance the case for the
ground state of the Bose-Hubbard model. Our results are valid for
such cases as well.

Expanding the atomic states in the number states basis
$\ket{n_0,n_1,\ldots,n_{M-1}}$, where the operators $\hat G$ and
$\hat Q$ are diagonal, we then find for the intensity of the
Bragg-reflected field
\begin{eqnarray}
\frac{\langle \hat n_{-k}(t)\rangle}{n_{\rm tot}}&=&
\frac{1}{2}-\frac{1}{2}\sum_{n_0,n_1,\ldots,n_{M-1}}P_{n_0,n_1,\ldots,n_{M-1}}
\nonumber \\
&\times&  \cos 2\omega(n_0,n_1,\ldots,n_{M-1})t,
\label{reflintensityformula}
\end{eqnarray}
where $P_{n_0,n_1,\ldots,n_{M-1}}$ is the probability to find
$n_m$ atoms in the $m^{\rm th}$ well and
$\omega(n_0,n_1,\ldots,n_{M-1})$ is the corresponding eigenvalue
of $\hat G$.

As we see in the following sections, this expression accounts
implicitly for the well-known property that Bragg scattering
depends strongly on the well separation $d$. In particular, the
light scattered from neighboring wells interferes either
constructively or destructively for well separations equal to
$\lambda/2$ or $\lambda/4$, respectively, two situations
relatively easy to realize experimentally.

\section{\label{twowellcase} Double-well potential}

Before considering the situation of a general optical lattice, we
discuss the simpler case of a double-well potential, where
explicit analytical results are readily obtained. Despite its
simplicity, this case already exhibits many of the properties of
the full $M$-site lattice and hence provides valuable intuition
for its understanding. We first discuss the Bragg-reflected
intensity for well separations $d=\lambda/2$ and $d=\lambda/4$,
and then turn to an arbitrary well separation and to the analysis
of the photon statistics.

\subsection{Reflected intensity}

\subsubsection{Well separation $d=\lambda/2$}

From Eq. (\ref{reflintensityformula}), most properties of the
reflected intensity can be understood from an analysis of the
eigenvalues of the operator $\hat G(d)$ describing the coupling
between the forward- and backward-propagating light fields. For
$d=\lambda/2$ we find from the definition (\ref{Nofd}) of
${\hat N}(d)$ that
$$
\hat G (\lambda/2)=g(\hat n_0+\hat n_1),
$$
indicating that the intensity of the Bragg-scattered light is only
sensitive to the total number of atoms $N$, a signature
of the constructive interference of the fields reflected off the
two wells. Specifically,
    \begin{equation}
    \frac{\langle \hat n_{-k}(t)\rangle}{n_{\rm tot}}=
    \sum_{N=0}^{\infty} P_N \sin^2gNt
    \label{nk1}
    \end{equation}
where $P_N$ is the probability that the total atom number in the two
wells is $N$.

\begin{figure}
\includegraphics{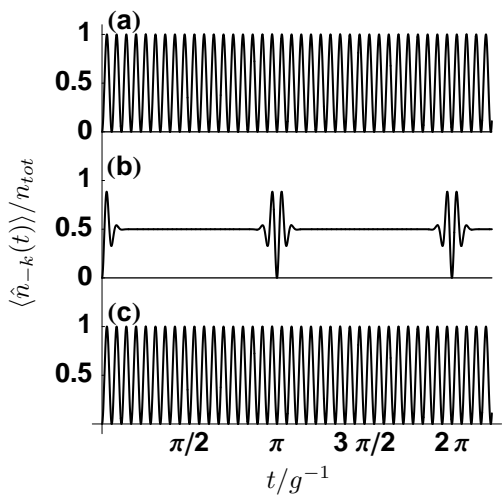}
\caption{Reflected intensities $\langle \hat n_{-k}(t)\rangle$ for well
separation $d=\lambda/2$ and for atoms (a) in a Fock state $\ket{\psi_{\rm
Mott}}$, (b) in a superfluid state in mean field approximation $\ket{\psi_{\rm
SF1}}$ and (c) in the number-conserving superfluid state $\ket{\psi_{\rm
SF2}}$. The mean number of atoms in each well is nine in all three cases.}
\label{nminusklambda2fig}
\end{figure}

Both the Fock state $\ket{\psi_{\rm Mott}}$ and superfluid state
$\ket{\psi_{\rm SF2}}$ have a well defined total number of atoms so
that the light field simply undergoes harmonic oscillations
between the $+k$ and $-k$ directions. These two situations are
illustrated in the upper and lower parts of Fig. \ref{nminusklambda2fig}.

In contrast, if the atomic field is described by coherent states
for each well, the total number of atoms is uncertain and
consequently the reflected intensity is comprised of oscillations
at the eigenfrequencies associated to all possible combinations of
atom numbers, leading to collapses and revivals similar to those
familiar from the two-photon Jaynes-Cummings model
\cite{twophotonJC1,twophotonJC2},
see the middle curve in Fig.
\ref{nminusklambda2fig}. The Bragg-reflected intensity
periodically collapses and revives after characteristic times
\begin{equation}
T^{\lambda/2,\rm SF1}_{\rm collapse}=\frac{1}{2g\Delta (\hat
n_0+\hat n_1)}
    =\frac{1}{2g\sqrt{\langle \hat n_0\rangle+\langle \hat n_1\rangle}},
    \label{Tcollapse}
\end{equation}
and
\begin{equation}
T_{\rm revival}=\pi/g \label{revivaltime},
\end{equation}
where $\Delta (\hat n_0+\hat n_1)=\sqrt{\langle(\hat n_0+\hat
n_1)^2\rangle-\langle\hat n_0+\hat n_1\rangle^2}$ is the standard
deviation of the total number of atoms. Equation (\ref{Tcollapse})
is important in that it indicates that the intensity of Bragg-scattered light is indeed sensitive to
fluctuations in the atom number. As such, it
offers a clear signature to distinguish the mean-field
description from the number-conserving description of the atomic
superfluid state.

\subsubsection{Well separation $d=\lambda/4$}

The light reflected from the two wells is now
$\pi$ out of phase and hence interferes destructively.
Mathematically, this is reflected by the coupling between the two
counter-propagating modes being now proportional to
the difference between the atom numbers in the two wells,
$$
\hat G(\lambda/4)=g|\hat n_0-\hat n_1|,
$$
so that the intensity of the backscattered light field is given by
    \begin{equation}
    \frac{\langle \hat n_{-k}(t)\rangle}{n_{\rm tot}}=
    \sum_{|\Delta N|=0}^{\infty} P_{|\Delta N|} \sin^2g|\Delta N|t,
    \label{refllightquarter}
    \end{equation}
where $P_{|\Delta N|}$ is the probability that the atomic
population difference between the two wells is $\Delta N$ or $-\Delta N$.

\begin{figure}
\includegraphics{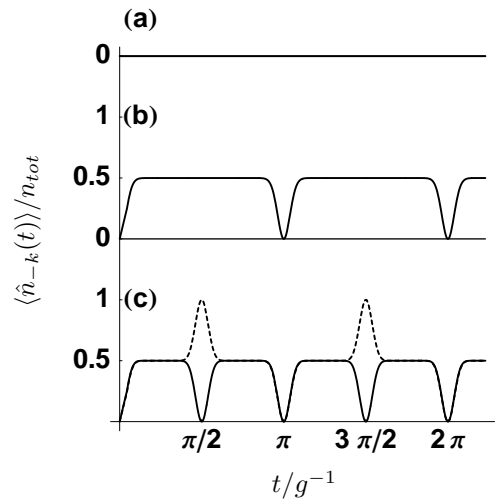}
\caption{Reflected intensities $\langle \hat n_{-k}(t)\rangle$ for
well separation $d=\lambda/4$ and for the atoms (a) in Fock states
with equal atom numbers, (b) in coherent states of equal mean atom
number and (c) in the number-conserving superfluid state. The mean
number of atoms in each well is nine in all cases. In (c) the solid line is for
$N=18$ and the dashed line is for $N=17$.}
\label{nminusklambda4fig}
\end{figure}

If the atomic fields in both wells are in Fock states, the sum in
Eq.~(\ref{refllightquarter}) reduces to just one term and the
light intensity undergoes sinusoidal oscillations at frequency
$g|n_0-n_1|$. If the atom numbers are equal the two optical modes
become decoupled, see Fig. \ref{nminusklambda4fig}(a).

If each well is in a coherent state, on the other hand, the number
difference is not known with certainty and accordingly the
oscillations of the intensities of the two modes contain frequency
components corresponding to all possible number differences.
In the case of equal mean atomic numbers in the two wells the
reflected intensity rises to half the initial intensity of the
forward-propagating field, remains constant at this value and
drops to zero at time $T_{\rm revival}$. The dips in the reflected
intensity can be thought of as `remnants' of the collapse and
revival features exhibited by the $d=\lambda/2$ case. Similarly to
that case the half-width of the dips is determined by the
uncertainty in the atom number difference, which is half of the
uncertainty in the total number of atoms \footnote{The standard
deviation is cut in half because negative frequencies do not
occur.}, so that the initial rise time is given by
$T^{\lambda/4,\rm SF1}_{\rm collapse}=2T^{\lambda/2,\rm SF1}_{\rm
collapse}=1/g\sqrt{\langle \hat n_0\rangle+\langle \hat
n_1\rangle}$.

Finally, for atoms in the number-conserving superfluid state
$\ket{\psi_{\rm SF 2}}$, odd differences in atomic numbers are
impossible for an even total number of atoms, hence all
possible oscillation frequencies differ by at least $2g$ (or $4g$
for the oscillations of the field intensities). For an even total
number of atoms the revival time is cut in half, see Fig.
\ref{nminusklambda4fig}(c). For odd numbers of atoms the reflected
intensity is equal to $n_{\rm tot}$ at times
$t=\pi/2g,\;3\pi/2g,\ldots$ Apart from these times the reflected
intensity behaves very similar to the coherent state case. The
collapse time is $T^{\lambda/4,\rm SF2}_{\rm
collapse}=2T^{\lambda/2,\rm SF1}_{\rm collapse}$.

\subsubsection{Arbitrary well separation}

The expressions (\ref{nk1}) and (\ref{refllightquarter}) for the
backscattered photon numbers in the two cases considered so far
indicate that the marginal probability distributions $P_N$ and
$P_{|\Delta N|}$ can be directly determined from the reflected
intensity by means of Fourier transforms. Unfortunately, the
complete number statistics $P_{n_0,n_1}$ for $n_0$ atoms in the
first well and $n_1$ atoms in the second well cannot be
reconstructed from these marginals.

The dynamics of the two optical modes becomes more involved for
arbitrary separations, as the eigenvalues of the operator $\hat G$
are no longer integer multiples of $g$ but in general irrational
numbers that depend on the occupation numbers $\hat n_0$ and $\hat
n_1$ in a nontrivial manner. One consequence is that the perfect
recurrences observed for $d=\lambda/2$ and $d=\lambda/4$ no longer
occur and the dynamics of the backscattered intensity is typically
non-periodic.

\begin{figure}
\includegraphics{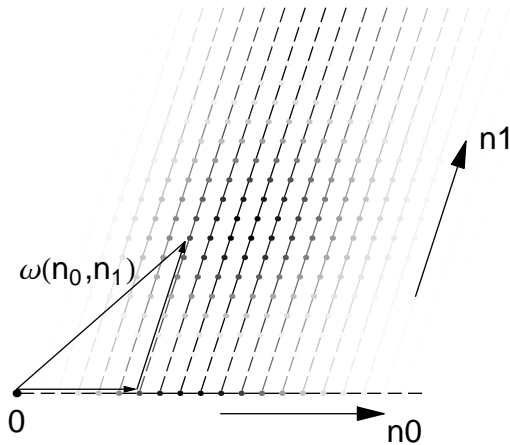}
\caption{Construction of the eigenvalues of $\hat G$ for
$d=\lambda/10$ and probability of occurrence of each frequency for
the two wells in coherent states with coherent amplitudes
$\alpha_1=\alpha_2=3$.} \label{twowellfrequenciesscheme}
\end{figure}

\begin{figure}
\includegraphics{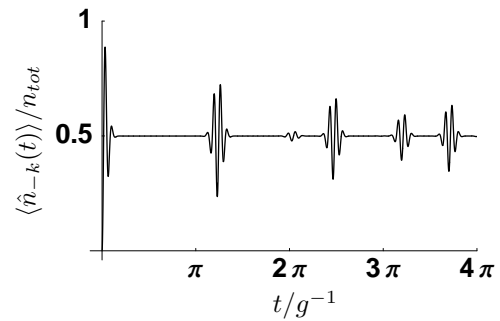}
\caption{Backscattered intensity for atoms in state
$\ket{\psi_{\rm SF1}}$ with nine atoms in each well on average and
a well separation $d=\lambda/10$.} \label{nmkgenerald}
\end{figure}

The backscattered intensity is determined by the spectral
properties of the operator $\hat G$ as we have seen, and these can
be understood by means of a simple geometric construction: For a
number state with $n_0$ atoms in the first well and $n_1$ atoms in
the second well the eigenvalue of $\hat G$ is given by the
hypothenuse of the triangle obtained by drawing a straight line of
length $g n_0$ and then continuing with another straight line of
length $g n_1$ at an angle $2kd$ with respect to the first line.
This construction is shown in Fig. \ref{twowellfrequenciesscheme},
which also illustrates the probability of each frequency occurring
if each well is in a coherent state with coherent amplitude
$\alpha_0=\alpha_1=3$. The dots along the $n_0$ line are drawn in
shades of gray with darker shades being more likely and lighter
shades being less likely. The $n_1$ lines originating from each
dot are drawn with the same convention, and the probability for a
given combination of $n_0$ and $n_1$, obtained by multiplying the
probabilities for $n_0$ and $n_1$, is again shown in gray scale.
Hence the most likely eigenfrequencies and the spread in
frequencies can be read off as the distances of the final dots
from the origin. A typical example of Bragg-reflected intensity is
shown in Fig. \ref{nmkgenerald}. This example is for nine atoms
per well in the coherent state $|\psi_{\rm SF1}\rangle$ and a well
separation $d=\lambda/10$.

\begin{figure}
\includegraphics{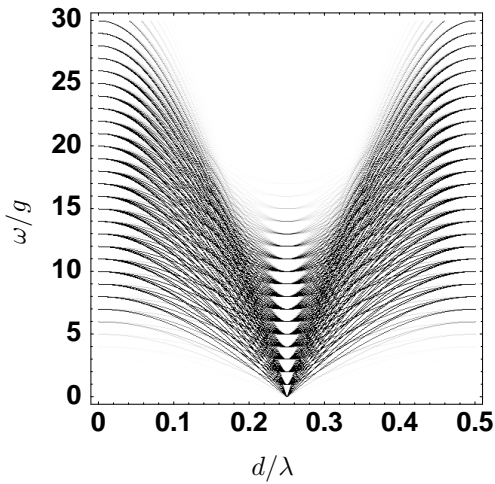}
\caption{Spectrum of $\hat G$ as a function of $d$. The darkness
of each eigenfrequency corresponds to its likelihood. In this
example both wells are in a coherent state with an average of 9
atoms per well.} \label{probomegad}
\end{figure}

Figure \ref{probomegad} shows the spectrum of $\hat G$ as a
function of $d$ for state $\ket{\psi_{\rm SF1}}$ with
$\alpha_0=\alpha_1=3$. Each frequency is again weighted with its
likelihood, and the eigenfrequencies have been collected in bins
of the size of the pixels in the figure. The special cases of
$d=\lambda/2$ and $d=\lambda/4$ are easily recognized, as for
those well separations only integer multiples of $g$ can occur.
For general well separations, the spectra lack any obvious
structure.

\begin{figure}
\includegraphics{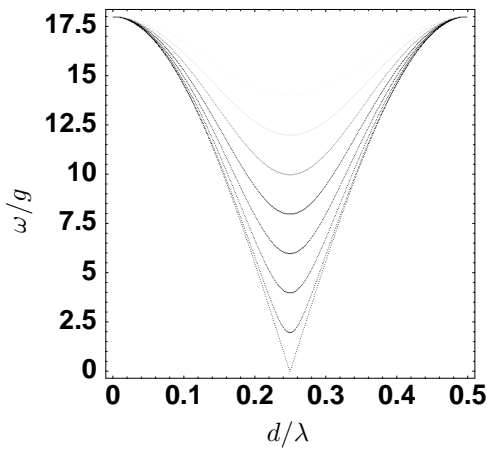}
\caption{Spectrum of $\hat G$ as a function of $d$ for atoms in state
    $\ket{\psi_{\rm SF2}}$ with 18 atoms.} \label{super2}
\end{figure}

The number-conserving superfluid state $|\psi_{\rm SF2}\rangle$
corresponds roughly to a ``diagonal cut'' through the construction
in Fig. \ref{twowellfrequenciesscheme} such that $n_0$ decreases
by one if $n_1$ is increased by one. Accordingly there are now
much fewer possible frequencies as can be seen in Fig.
\ref{super2}. Furthermore, the width of the frequency distribution
is no longer independent of $d$. We have $\Delta |\hat n_0-\hat
n_1|=\sqrt{\frac{N}{2}(1-\cos 2kd)}$
\footnote{\label{omegameanfoot}This result is valid for well
separations for which the mean frequency is larger than the
frequency fluctuations. If $d$ approaches $\lambda/4$ and the mean
frequency becomes comparable to the fluctuations the width of the
frequency distribution approaches the value for $d=\lambda/4$.}.

\subsection{Photon number statistics}

A major advantage of the two-well case over the full lattice case
is that we can easily solve the Schr\"odinger equation for the
coupled atom-cavity system numerically. This integration is
greatly facilitated by the conservation of the number of atoms in
each well and the total number of photons, which we exploit by
integrating in the invariant subspaces of constant $\hat n_0$,
$\hat n_1$ and $\hat n_k+\hat n_{-k}$. From the full solution we
can gain more detailed information about the dynamics of the
photons than the average reflected intensity considered so far.

\begin{figure}
\includegraphics{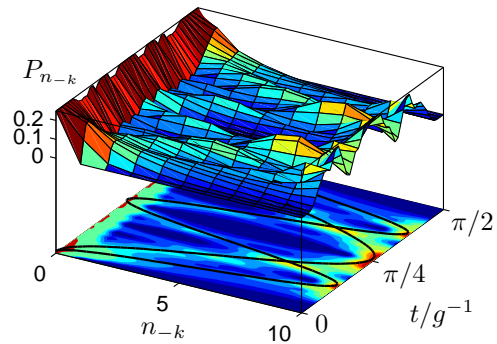}
\caption{(Color online) Photon number statistics for atoms in
state $\ket{\psi_{\rm SF2}}$ for $N=18$ atoms and a well
separation $d=\lambda/4$. Also indicated are the elementary
sinusoidal oscillations corresponding to number differences one
and two.} \label{numberstatsf2}
\end{figure}

Here we consider the number statistics of the Bragg-reflected photons,
\begin{equation}
P_{n_{-k}}=\langle \ket{n_{-k}}\bra{n_{-k}}\rangle.
\end{equation}
The calculation of $P_{n_{-k}}$ involves a trace over the well
occupation numbers. As a consequence the number statistics is, in
complete analogy to the reflected intensity in Eq.
(\ref{reflintensityformula}), the sum of the number statistics
$P_{n_{-k}}^{(n_0,n_1)}$ for each set of occupation numbers
$(n_0,n_1)$, each weighted with its probability of occurring in
the atomic state,
$$
P_{n_{-k}}=\sum_{n_0,n_1} P_{n_0,n_1}P_{n_{-k}}^{(n_0,n_1)}.
$$
\begin{figure}
\includegraphics{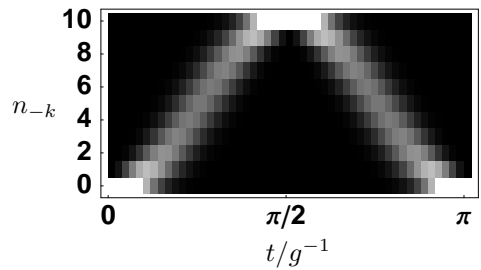}
\caption{Photon number statistics $P_{n_{-k}}(t)$ for each well in a Fock
state with $n_0=9$ and $n_1=10$ atoms, a well separation of
$d=\lambda/4$ and 10 photons initially in the $+k$ direction.
Lighter shades of grey correspond to higher probabilities.}
\label{numberstatFock}
\end{figure}
Figure \ref{numberstatFock} shows one of the building blocks $P_{n_{-k}}^{(n_0,n_1)}(t)$ for
well occupation numbers $n_0=10$ and $n_1=9$ with $d=\lambda/4$.
In this case the two counter-propagating modes are linearly
coupled to each other, with coupling frequency $g|n_0-n_1|$.
Figure \ref{numberstatsf2} shows the number statistics for atoms
in state $\ket{\psi_{\rm SF2}}$.
This example is representative of the typical situation, with
non-trivial photon statistics and `complicated' dynamics of the
light field that cannot be adequately described by means of the
reflected intensity only. As the elementary probability
wavepackets $P_{n_{-k}}^{(n_0,n_1)}(t)$ dephase with respect to each other the
photon statistics evolves through complex patterns, with a large uncertainty of
the photon number. The reflected light intensity
exhibits large fluctuations the description of which requires all
moments of the intensity up to order $n_{\rm tot}$.

\begin{figure}
\includegraphics{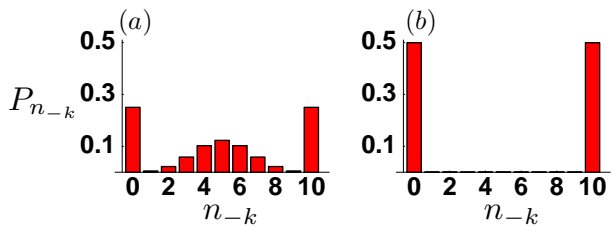}
\caption{(Color online) Photon number statistics at times (a)
$t=\pi/4 g^{-1}$ and (b) $t=\pi/2 g^{-1}$ for atoms in state
$\ket{\psi_{\rm SF1}}$ with average atom numbers $n_0=n_1=9$ for
well separation $d=\lambda/4$.} \label{photonstatpi2}
\end{figure}

There are however special times when several of the elementary
photon statistics building blocks add up to yield peaks at specific
photon numbers. Prominent such times are $t=\pi/2 g^{-1}$ and
$t=\pi/4 g^{-1}$ for $d=\lambda/4$. At these times all odd
frequency components (i.e. odd number differences) lead to $\hat
n_{-k}=n_{\rm tot}$ while all even number differences lead to
$\hat n_{-k}=0$. This is illustrated in Fig. \ref{photonstatpi2}
for the atomic state $\ket{\psi_{\rm SF1}}$. For the fairly large
mean number of atoms per well considered in that example the
probabilities for even and odd frequencies are approximately equal
so that the probabilities to find all photons reflected or no
reflection at all are also nearly equal. For atoms in state
$\ket{\psi_{\rm SF2}}$ the photon statistics at time $t=\pi/4
g^{-1}$ look very similar to Fig. \ref{photonstatpi2}(b).
It is important to emphasize that the light field does not evolve
into a `NOON' state $(\ket{n_{\rm tot},0}+\ket{0,n_{\rm
tot}})/\sqrt{2}$, which would show the same behavior. Rather, the
light field is entangled with the atoms in the Schr\"odinger cat
state
\begin{widetext}
\begin{eqnarray}
&&\ket{\psi(t=\pi/2g)}=\ket{\psi_{\rm SC}} \nonumber \\
    &=&\frac{1}{2}\ket{n_{\rm tot}, 0}
    \left [\ket{\alpha_0,(-1)^{n_{\rm tot}}\alpha_1}
    +\ket{-\alpha_0,-(-1)^{n_{\rm tot}}\alpha_1} \right ]
    +\frac{(-i)^{n_{\rm tot}}}{2}\ket{0,n_{\rm tot}}\left [\ket{\alpha_0,(-1)^{n_{\rm tot}}\alpha_1}
    -\ket{-\alpha_0,-(-1)^{n_{\rm tot}}\alpha_1}\right ] \nonumber
\end{eqnarray}
\end{widetext}
which is reflected by the atomic $Q$-function in the first well,
$$
Q(\alpha)=\langle\psi_{\rm SC}|\alpha\rangle\langle\alpha
\ket{\psi_{\rm SC}}\nonumber\\
=\frac{1}{2\pi}\left(e^{-|\alpha-\alpha_0|^2}+e^{-|\alpha+\alpha_0|^2}\right).
$$

For $d=\lambda/2$ the number statistics looks qualitatively
similar to the $\lambda/4$ case. The most likely frequencies are
however typically higher because of the constructive interference
of the light reflected off the two wells that leads to the
coupling frequency being $g(\hat n_0+\hat n_1)$.

In practice, Fock states are very hard to realize. Typically the
$+k$ mode will start out in a superposition of states with
different photon numbers such as a coherent state. As we have mentioned, this
case is qualitatively similar to the Fock state case and here we would
like to explain in more detail why this is so. Because the system
Hamiltonian conserves the total number of photons $\hat
n_{-k}+\hat n_{k}$, no interference between states with different
total photon numbers can arise in any observable that preserves
the total photon number as well. The reflected intensity and the
photon number statistics are clearly observables of this type. For
a general state the expectation value of these observables is the
sum of the expectation values for the various photon numbers
weighted with the probability of that photon number. The
consequence is simply that one has additional fluctuations in the
reflected intensity due to the uncertainty of the photon number in
the initial state. Another way to see that the intensity behaves
the same for all initial states of the $+k$ mode is to insert the
formal solution for the modes of the light field Eq.
(\ref{formalsolution}) into the expectation value of the
intensity. The terms containing operators of the $-k$ mode vanish
and we end up with the expectation value of the intensity in the
$+k$ mode at time $t=0$ multiplying the trace of the atomic
operators over the atomic states.

\section{\label{latticecase}Optical Lattice}

\subsection{Well separation $d=\lambda/2$}

This section extends the results of the double well analysis to
the case of an optical lattice with a large number of wells. The
case of a well separation $d=\lambda/2$ is again particularly
simple since in that case $\hat G$ is simply proportional to the
total number of atoms in the lattice,
\begin{equation}
\hat G=g\sum_m \hat n_m\equiv g\hat N.
\end{equation}
Both the Fock state and the number-conserving superfluid state
$\ket{\psi_{\rm SF 2}}$ have a well-defined total number of atoms
$N$, hence the intensity of the optical field undergoes simple
sinusoidal oscillations at frequency $g N$ between the two
counter-propagating modes.

The situation is more complicated in the mean-field description of
the superfluid state, in which case the matter-wave field at each
lattice site is in a coherent state with mean number of atoms
$n_m=\langle \hat N\rangle/M$. In the limit
of a large lattice, $M \gg 1$, the central limit theorem permits
to approximate the probability distribution of the total number of
atoms to an excellent degree as
\begin{equation}
P_N=\frac{1}{\sqrt{2\pi \langle
\hat N\rangle}}\exp\left[{-\frac{(N-\langle \hat N\rangle)^2}{2\langle
\hat N\rangle}}\right].
\end{equation}
In the limit of large $\langle \hat N \rangle$ this distribution
is sharply peaked at $\langle \hat N \rangle$ and the dynamics of
the light is largely characterized by oscillations at frequency $g
\langle \hat N \rangle$ between the two counter-propagating modes.
However the finite variance of the total atomic number
distribution results in a collapse of these oscillations on a time
scale
\begin{equation}
T^{\lambda/2,{\rm SF1},latt.}_{\rm collapse}=\frac{1}{2g\sqrt{\langle
\hat N\rangle}}.
\end{equation}
The photons undergo roughly $2\sqrt{\langle \hat N \rangle}$
oscillations before this collapse. Note that since the allowed
frequencies are discrete and integer multiples of $g$, the
oscillations also show complete revivals, with a revival time
given by Eq. (\ref{revivaltime}).

\subsection{Well separation $d=\lambda/4$}

The situation is markedly different for a well separation
$d=\lambda/4$. In that case $\hat G$ is proportional to the
difference in the total number of atoms trapped on even and odd
lattice sites,
\begin{equation}
\hat G=g\left|\sum_{m\; {\rm even}} \hat n_m-\sum_{m\; {\rm odd}}
\hat n_m\right|\equiv g|\hat N_e -\hat N_o|,
    \label{grefl4}
\end{equation}
and the light field probes fluctuations in the occupation
differences of neighboring sites. This is a useful property since
fluctuations of this type, while being typically difficult to
measure, can serve to distinguish various many-body quantum states
such as the Mott insulator state and the superfluid state. Note
that if all lattice sites are in Fock states with equal atomic
population, then $\hat G$ vanishes identically for an even number
of lattice sites and is very small when the number of lattice
sites is odd. In typical experiments the atoms are also subject to
a harmonic trap and the added trapping potential leads to shells
with constant occupation number separated from each other by edges
with a superfluid component. Equation (\ref{grefl4}) shows that
the reflected light can serve as a good detector of these edges.

For the atoms in the mean-field superfluid state $\ket{\psi_{\rm
SF 1}}$ the coupling strength can again be analyzed using the
central limit theorem. To this end we use the fact that $N_e$
and $-N_o$ are the sums of a large number of independent
random variables with means $\langle \hat N\rangle/2$ and $-\langle \hat N
\rangle/2$, respectively, and standard deviation $\sqrt{\langle
\hat N\rangle/2}$. From the resulting Gaussian probability
distributions for $ N_e$ and $N_o$ we readily find the
probability distribution for $N_e-N_o$ by using that the
sum of two Gaussian random variables is again a Gaussian. For the
intensity of the reflected light it is actually $|N_e-
N_o|$ that is important, see Eq. (\ref{refllightquarter}). The
probability distribution for this quantity is readily found from
those for $N_{e,o }$ as
\begin{equation}
P_{|N_e-N_o|}= (2-\delta_{N_e,N_o}) \frac{1}{\sqrt{2\pi
\langle \hat N\rangle}}\exp\left[{-\frac{(N_e-N_o)^2}{2\langle
\hat N\rangle}}\right],
\end{equation}
where $\delta$ is the Kronecker delta.

The analysis of the state $\ket{\psi_{\rm SF 2}}$ requires a bit more
attention. Since $\hat N_e$ is the sum of many operators it is
natural to assume that its distribution is Gaussian, and hence
fully characterized by its mean $\langle \psi_{\rm SF 2}|\hat
N_e|\psi_{\rm SF2}\rangle=N  /2$ and standard
deviation $\Delta N_e=\sqrt{ N }/2$. Since every
choice of $N_e$ automatically determines $N_o=N-N_e$ the
probability distribution for $|N_e-N_o|$ is
\begin{eqnarray}
P_{|N_e-N_o|}&=&\delta_{N_o+N_e,N}(2-\delta_{N_e,N_o})\nonumber\\
    &\times&\frac{\sqrt{2}}{\sqrt{\pi  N }}
    \exp\left[-\frac{(N_e-N_o)^2}{2 N }\right].
    \label{Pnl4sf2}
\end{eqnarray}

\begin{figure}
\includegraphics{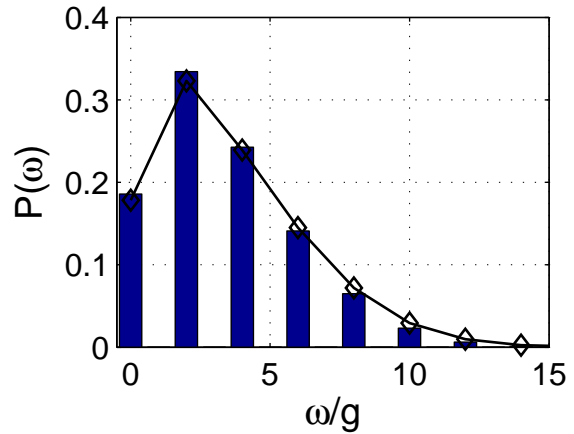}
\caption{(Color online) Probability distribution of the coupling
frequencies (eigenvalues of ${\hat G}$ for $d=\lambda/4$ with
atoms in state $\ket{\psi_{\rm SF2}}$ for 10 lattice sites with 20
atoms. The bars are the exact probabilities determined numerically
and the diamonds show the distribution obtained from Eq.
(\ref{Pnl4sf2}). The line is an aid to the eye.}
\label{spectr_l4sf2}
\end{figure}

Figure \ref{spectr_l4sf2} shows an example of the numerically
obtained exact probability distribution and compares it to the
approximate result of Eq. (\ref{Pnl4sf2}) for as few as 10 lattice
sites and 20 atoms. Clearly, the two results already agree quite
well even for these modest numbers, so
that the expression (\ref{Pnl4sf2}) can be considered as
essentially exact for the much larger numbers of lattice sites and
atoms typically encountered in practice.

Note that if one assumes that the occupation numbers of the
lattice sites contributing to $\hat N_e$ are independent when
calculating the probability distribution for $N_e$ with the
central limit theorem, one finds that the standard deviation of
the resulting distribution is too large by a factor of $\sqrt{2}$.
In this sense, fluctuations of the number difference between even
and odd sites are suppressed in the state $\ket{\psi_{\rm SF2}}$.

\subsection{General well separation $d$}

The calculation of the possible eigenvalues of $\hat G$ and their
probabilities of occurring for the atomic states $\ket{\psi_{\rm
SF1}}$ and $\ket{\psi_{\rm SF2}}$ by means of the central limit
theorem can readily be extended to well separations
$d=q\lambda/2p$, where $q$ and $p$ are integers. The lattice sites
can be grouped into $p$ distinct classes according to the roots of
unity $e^{2ikd}=e^{2i\pi l/p}$ with $l=0,1,\ldots,p-1$. The total
number of atoms in each class \footnote{For simplicity we have
assumed that $M$ is a multiple of $p$. The extension to more
general numbers of lattice sites is straightforward.},
    $$
    \hat N_l=\sum_{m=0}^{M/p-1}\hat n_{l+mp}
    $$
enters the frequency with the common coefficient $e^{2i\pi l/p}$
and each class comprises many lattice sites, so that the central
limit theorem can be used to calculate the associated number
distribution $P_{N_l}$ \footnote{As we have mentioned above the
$\hat N_l$ are not independent for $p=2$, but we have
numerically verified their independence for $p>2$.}. The
frequencies are then obtained as a function of the few macroscopic
occupation numbers $\hat N_l$, which are independent to a good
approximation. Thus the probabilities for the frequencies
$\omega(N_0,N_1,\ldots,N_{p-1})=g|\sum_{l=0}^{p-1}N_le^{2i\pi
l/p}|$ are simply the product of the probabilities for each
macroscopic occupation $P_{N_l}$. These probabilities can however
not be written down in closed form, since the frequencies are no
longer sums of Gaussian variables.

\begin{figure}
\includegraphics{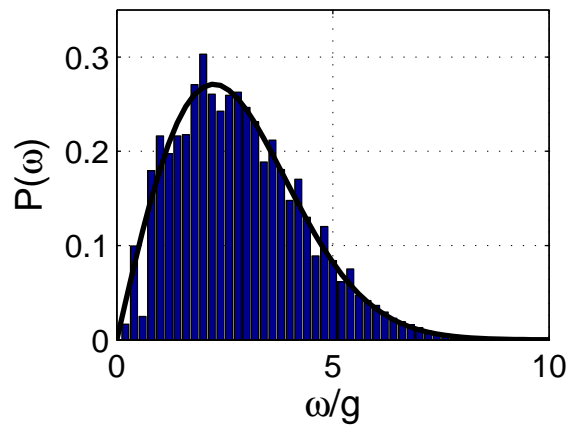}
\caption{(Color online) Probability distribution of frequencies
for atoms in the mean-field superfluid state $\ket{\psi_{\rm
SF1}}$ with ten lattice sites and ten atoms for $d=\lambda/10$.
The solid line is the distribution Eq. (\ref{rwdistribution}), and
the histogram the result of an exact numerical diagonalization of
${\hat G}(d)$.} \label{pomegasf1}
\end{figure}

\begin{figure}
\includegraphics{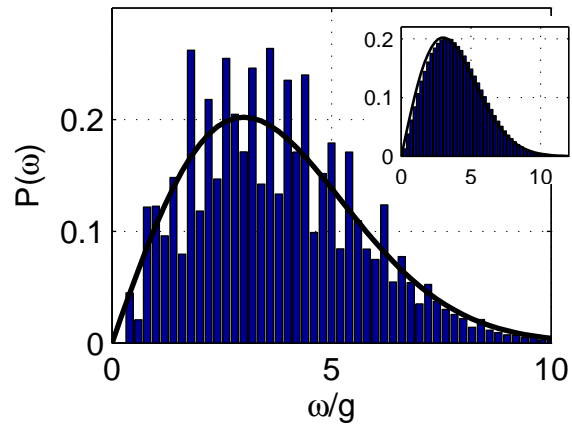}
\caption{(Color online) Probability distribution of frequencies
for atoms in the number-conserving superfluid state
$\ket{\psi_{\rm SF2}}$ with ten lattice sites and 18 atoms for
$d=\lambda/10$. The solid line is the distribution Eq.
(\ref{rwdistribution}) and the histogram a result of the numerical
diagonalization of ${\hat G}(d)$. The inset shows the same
distribution for $d=\frac{\sqrt{2}}{10}\lambda$.}
\label{pomegasf2}
\end{figure}

If $p$ is large or $d$ is an irrational fraction of $\lambda$, the
behavior of the system becomes identical for atoms in states
$\ket{\psi_{\rm SF1}}$ and $\ket{\psi_{\rm SF2}}$ in the limit of
a large number of lattice sites. In this case we can think of
$\hat N(d)$ as the result of a random walk of $M$ steps of average
stride length $\langle N \rangle/M$ in the complex plane and the
frequency is proportional to the distance from the origin to the
final point. The distribution of possible frequencies $\omega$
becomes quasi-continuous with
\begin{equation}
P(\omega)=\frac{2\omega}{g^2\langle \hat N
\rangle}\exp\left[-\frac{\omega^2}{g^2\langle \hat N \rangle}\right],
\label{rwdistribution}
\end{equation}
independently of whether the atoms are in state $\ket{\psi_{\rm
SF1}}$ or $\ket{\psi_{\rm SF2}}$. This is illustrated in Fig.
\ref{pomegasf1} for atoms in state $\ket{\psi_{\rm SF 1}}$ and in
Fig. \ref{pomegasf2} for atoms in state $\ket{\psi_{\rm SF2}}$. It
is remarkable that the numerically obtained true frequency
distribution is in a good agreement with the approximate Eq.
(\ref{rwdistribution}) even for the low number of atoms and
lattice sites considered here, and even more so, that this
agreement is already obtained for $p=5$. As one might expect, we
find much better agreement between the actual frequency
distributions and Eq. (\ref{rwdistribution}) for the atom and site
numbers of Figs. \ref{pomegasf1} and \ref{pomegasf2} if the
lattice spacings correspond either to a larger $p$ or to
irrational fractions of the optical wavelength, as is the case in
the inset of Fig. \ref{pomegasf2}, which is for
$d=\sqrt{2}\lambda/10$.

One can calculate the reflected intensity in closed form from the
frequency distribution (\ref{rwdistribution}) by approximating the
sum in Eq. (\ref{reflintensityformula}) by an integral,
\begin{eqnarray}
\frac{\langle\hat n_{-k}\rangle}{n_{\rm tot}}&=&\int_0^\infty
d\omega P(\omega)\sin^2\omega t\nonumber\\
    &=&\frac{g\langle\hat N\rangle\sqrt{\pi}}{2\sqrt{M}}|t|e^{-\frac{g^2\langle\hat N\rangle^2t^2}{M}}
    {\rm erfi}\left(\frac{g\langle \hat N\rangle t}{\sqrt{M}}\right),
\end{eqnarray}
where ${\rm erfi}(x)=\frac{2}{i\sqrt{\pi}}\int_0^{ix}dze^{-z^2}$
is the complex error function. The intensity converges to $\langle
\hat n_{-k}\rangle=n_{\rm tot}/2$ after a transient of
duration $T_{\rm collapse}\approx \sqrt{M}/(g\langle \hat N \rangle)$.

\begin{table}
\renewcommand{\arraystretch}{1}
\begin{tabular}{|c|c||c|c|}
\hline
lattice&atomic&\multicolumn{2}{c|}{$T_{\rm collapse}^{-1}$}\\
\cline{3-4}
spacing&state&2 wells&lattice\\
\hline
\hline
&$\ket{\psi_{\rm Mott}}$&$0$&$0$\\
\cline{2-4}
$d=\frac{\lambda}{2}$&$\ket{\psi_{\rm SF1}}$&$2g\sqrt{\langle
\hat n_0\rangle+\langle\hat n_1\rangle}$&$2g \sqrt{\langle\hat N\rangle}$\\
\cline{2-4}
&$\ket{\psi_{\rm SF2}}$&$0$&$0$\\
\hline
&$\ket{\psi_{\rm Mott}}$&$0$&$0$\\
\cline{2-4}
$d=\frac{\lambda}{4}$&$\ket{\psi_{\rm SF1}}$&$g\sqrt{\langle
\hat n_0\rangle+\langle\hat n_1\rangle}$&$g \sqrt{1-2/\pi}\sqrt{\langle \hat N\rangle}$\\
\cline{2-4}
&$\ket{\psi_{\rm SF2}}$&$g\sqrt{N}$&$g\sqrt{1-2/\pi}\sqrt{N}$\\
\hline
&$\ket{\psi_{\rm Mott}}$&$0$&$0$\\
\cline{2-4}
general $d$&$\ket{\psi_{\rm SF1}}$&$2g\sqrt{\langle
\hat n_0\rangle+\langle \hat n_1 \rangle}$
&$\sqrt{1-\pi/4}g\langle \hat N\rangle/\sqrt{M}$\\
\cline{2-4}
&$\ket{\psi_{\rm SF2}}$&$2g\sqrt{\frac{N}{2}
(1-\cos 2kd)}$&$\sqrt{1-\pi/4}gN/\sqrt{M}$\\
\hline
\end{tabular}
\caption{Collapse times for the various lattice spacings and
atomic states. See the comments in footnote [21] regarding the
validity of the double well results for $\ket{\psi_{\rm SF1}}$ and
$\ket{\psi_{\rm SF2}}$.} \label{collapsetimestabel}
\end{table}

\section{\label{conclusion}Conclusion}

In this paper we have investigated the Bragg reflection of a
quantized light field off ultracold atoms trapped in an optical
lattice as a function of the lattice spacing. We have considered
atoms in a Mott state as well as a superfluid state, the latter
described both in the mean-field approximation and in a
number-conserving form. We have studied both a simplified two-mode
model that allows us to carry out most of the calculations explicitly
or numerically as well as the full lattice case, where our
analysis mostly relies on statistical arguments.

Our results show that the dynamics of the light field strongly
depends on the many-particle state of the matter-wave field and
thus can serve as a diagnostic tool for that state. Depending on
the lattice spacing the light field can probe fluctuations of
various density correlation functions on the lattice. These
fluctuations typically lead to collapses of the oscillations of
the reflected intensity. From the standpoint of probing the atomic
many-particle state the lattice spacing $d=\lambda/4$ seems the
most promising, since in that case the three atomic states give
rise to dramatically different reflection signals: The Mott state
leads to perfect sinusoidal oscillations, the superfluid state
described in the mean-field approximation to collapses with
revivals after $t=\pi/g$ and the number-conserving superfluid
state to revivals after $t=\pi/2g^{-1}$ for even total atom
numbers and anti-revivals at the same times for odd total atom
numbers. Also, the sensitivity of the reflected intensity to the
number differences on neighboring sites means that it can be used
as a detector for the edges of the density plateaus in the Mott
insulator state in a harmonic trapping potential.

The double-well case can easily be handled by numerically solving
for the coupled dynamics of atoms and light field. The solution
shows that the atoms and light field evolve into non-trivial
entangled states if the atoms are in a superfluid state. For those
states the photon number statistics is typically many-modal and
the characterization of the state of the light field merely in
terms of the reflected intensity is incomplete.

Most results obtained for the double-well case immediately carry
over to the case of a lattice, where we have shown that for
general well separations, the system behaves the same for the
mean-field and the number-conserving superfluid state.

Future work will extend this model to include cavity losses
thereby making it possible to model more realistically the
measurement of the reflected photons. It will be interesting to
determine into which atomic states the matter-wave field is
projected as a result of this measurement. Another open question
is the determination of the best possible choice of lattice period
$d$, or even of an optimal sequence of lattice spacings for the
purpose of reconstructing the full counting statistics of the
atomic field. We plan to investigate this point in more detail by
studying the information content of the reflection signal by means
of a Bayseian analysis.

\section*{Acknowledgements}

While preparing this manuscript we became aware of closely related
research being carried out in the group of H. Ritsch
\cite{Mekhov:ProbingLattice}. We thank Dr. I. B. Mekhov for stimulating
discussions.  This work is supported in part by the US Office of Naval
Research, by the National Science Foundation, by the US Army Research Office,
by the Joint Services Optics Program, and by the National Aeronautics and Space
Administration.

\bibliography{mybibliography}

\begin{thebibliography}{21}
\expandafter\ifx\csname natexlab\endcsname\relax\def\natexlab#1{#1}\fi
\expandafter\ifx\csname bibnamefont\endcsname\relax
  \def\bibnamefont#1{#1}\fi
\expandafter\ifx\csname bibfnamefont\endcsname\relax
  \def\bibfnamefont#1{#1}\fi
\expandafter\ifx\csname citenamefont\endcsname\relax
  \def\citenamefont#1{#1}\fi
\expandafter\ifx\csname url\endcsname\relax
  \def\url#1{\texttt{#1}}\fi
\expandafter\ifx\csname urlprefix\endcsname\relax\def\urlprefix{URL }\fi
\providecommand{\bibinfo}[2]{#2}
\providecommand{\eprint}[2][]{\url{#2}}

\bibitem[{lew()}]{lewensteincondmat}
\bibinfo{note}{See the pre-print by M. Lewenstein et. al., arXiv:
  cond-mat/0606771v1 (2006) for a recent review}.

\bibitem[{\citenamefont{Jaksch et~al.}(1998)\citenamefont{Jaksch, Bruder,
  Cirac, and Zoller}}]{Jaksch:BECinLattice}
\bibinfo{author}{\bibfnamefont{D.}~\bibnamefont{Jaksch}},
  \bibinfo{author}{\bibfnamefont{C.}~\bibnamefont{Bruder}},
  \bibinfo{author}{\bibfnamefont{J.~I.} \bibnamefont{Cirac}}, \bibnamefont{and}
  \bibinfo{author}{\bibfnamefont{P.}~\bibnamefont{Zoller}},
  \bibinfo{journal}{Phys. Rev. A} \textbf{\bibinfo{volume}{81}},
  \bibinfo{pages}{3108} (\bibinfo{year}{1998}).

\bibitem[{\citenamefont{Jaksch and Zoller}(2005)}]{Zoller:OLreview}
\bibinfo{author}{\bibfnamefont{D.}~\bibnamefont{Jaksch}} \bibnamefont{and}
  \bibinfo{author}{\bibfnamefont{P.}~\bibnamefont{Zoller}},
  \bibinfo{journal}{Ann. Phys.} \textbf{\bibinfo{volume}{315}},
  \bibinfo{pages}{52} (\bibinfo{year}{2005}).

\bibitem[{\citenamefont{{J. J. Garcia-Ripoll} et~al.}(2004)\citenamefont{{J. J.
  Garcia-Ripoll}, {M. A. Martin-Delgado}, and {J. I.
  Cirac}}}]{GarciaRipoll:SpinHamiltonians}
\bibinfo{author}{\bibnamefont{{J. J. Garcia-Ripoll}}},
  \bibinfo{author}{\bibnamefont{{M. A. Martin-Delgado}}}, \bibnamefont{and}
  \bibinfo{author}{\bibnamefont{{J. I. Cirac}}}, \bibinfo{journal}{Phys. Rev.
  Lett.} \textbf{\bibinfo{volume}{93}}, \bibinfo{pages}{250405}
  (\bibinfo{year}{2004}).

\bibitem[{\citenamefont{Miyakawa and Meystre}(2006)}]{Miyakawa:AndersonModel}
\bibinfo{author}{\bibfnamefont{T.}~\bibnamefont{Miyakawa}} \bibnamefont{and}
  \bibinfo{author}{\bibfnamefont{P.}~\bibnamefont{Meystre}},
  \bibinfo{journal}{Phys. Rev. A} \textbf{\bibinfo{volume}{73}},
  \bibinfo{pages}{021601(R)} (\bibinfo{year}{2006}).

\bibitem[{\citenamefont{Greiner
  et~al.}(2002{\natexlab{a}})\citenamefont{Greiner, Mandel, Esslinger,
  H\"ansch, and Bloch}}]{Bloch:MottInsulator1}
\bibinfo{author}{\bibfnamefont{M.}~\bibnamefont{Greiner}},
  \bibinfo{author}{\bibfnamefont{O.}~\bibnamefont{Mandel}},
  \bibinfo{author}{\bibfnamefont{T.}~\bibnamefont{Esslinger}},
  \bibinfo{author}{\bibfnamefont{T.~W.} \bibnamefont{H\"ansch}},
  \bibnamefont{and} \bibinfo{author}{\bibfnamefont{I.}~\bibnamefont{Bloch}},
  \bibinfo{journal}{Nature (London)} \textbf{\bibinfo{volume}{415}},
  \bibinfo{pages}{39} (\bibinfo{year}{2002}{\natexlab{a}}).

\bibitem[{\citenamefont{Greiner
  et~al.}(2002{\natexlab{b}})\citenamefont{Greiner, Mandel, H\"ansch, and
  Bloch}}]{Bloch:MottInsulator2}
\bibinfo{author}{\bibfnamefont{M.}~\bibnamefont{Greiner}},
  \bibinfo{author}{\bibfnamefont{O.}~\bibnamefont{Mandel}},
  \bibinfo{author}{\bibfnamefont{T.~W.} \bibnamefont{H\"ansch}},
  \bibnamefont{and} \bibinfo{author}{\bibfnamefont{I.}~\bibnamefont{Bloch}},
  \bibinfo{journal}{Nature (London)} \textbf{\bibinfo{volume}{419}},
  \bibinfo{pages}{51} (\bibinfo{year}{2002}{\natexlab{b}}).

\bibitem[{\citenamefont{Rom et~al.}(2004)\citenamefont{Rom, Best, Mandel,
  Widera, Greiner, {T. W. H\"ansch}, and Bloch}}]{Rom:MoleculesLattices}
\bibinfo{author}{\bibfnamefont{T.}~\bibnamefont{Rom}},
  \bibinfo{author}{\bibfnamefont{T.}~\bibnamefont{Best}},
  \bibinfo{author}{\bibfnamefont{O.}~\bibnamefont{Mandel}},
  \bibinfo{author}{\bibfnamefont{A.}~\bibnamefont{Widera}},
  \bibinfo{author}{\bibfnamefont{M.}~\bibnamefont{Greiner}},
  \bibinfo{author}{\bibnamefont{{T. W. H\"ansch}}}, \bibnamefont{and}
  \bibinfo{author}{\bibfnamefont{I.}~\bibnamefont{Bloch}},
  \bibinfo{journal}{Phys. Rev. Lett.} \textbf{\bibinfo{volume}{93}},
  \bibinfo{pages}{073002} (\bibinfo{year}{2004}).

\bibitem[{\citenamefont{{T. St\"oferle} et~al.}(2006)\citenamefont{{T.
  St\"oferle}, Moritz, {K. G\"unther}, {M. K\"ohl}, and {T.
  Esslinger}}}]{Stoeferle:MoleculesLattice}
\bibinfo{author}{\bibnamefont{{T. St\"oferle}}},
  \bibinfo{author}{\bibfnamefont{H.}~\bibnamefont{Moritz}},
  \bibinfo{author}{\bibnamefont{{K. G\"unther}}},
  \bibinfo{author}{\bibnamefont{{M. K\"ohl}}}, \bibnamefont{and}
  \bibinfo{author}{\bibnamefont{{T. Esslinger}}}, \bibinfo{journal}{Phys. Rev.
  Lett.} \textbf{\bibinfo{volume}{96}}, \bibinfo{pages}{030401}
  (\bibinfo{year}{2006}).

\bibitem[{\citenamefont{Thalhammer et~al.}(2006)\citenamefont{Thalhammer,
  Winkler, Lang, Schmid, Grimm, and {J. Hecker
  Denschlag}}}]{Thalhammer:MoleculesLattice}
\bibinfo{author}{\bibfnamefont{G.}~\bibnamefont{Thalhammer}},
  \bibinfo{author}{\bibfnamefont{K.}~\bibnamefont{Winkler}},
  \bibinfo{author}{\bibfnamefont{F.}~\bibnamefont{Lang}},
  \bibinfo{author}{\bibfnamefont{S.}~\bibnamefont{Schmid}},
  \bibinfo{author}{\bibfnamefont{R.}~\bibnamefont{Grimm}}, \bibnamefont{and}
  \bibinfo{author}{\bibnamefont{{J. Hecker Denschlag}}},
  \bibinfo{journal}{Phys. Rev. Lett.} \textbf{\bibinfo{volume}{96}},
  \bibinfo{pages}{050402} (\bibinfo{year}{2006}).

\bibitem[{\citenamefont{Jaksch et~al.}(2002)\citenamefont{Jaksch, Venturi,
  Cirac, Williams, and Zoller}}]{Jaksch:creationmoleculeinOL}
\bibinfo{author}{\bibfnamefont{D.}~\bibnamefont{Jaksch}},
  \bibinfo{author}{\bibfnamefont{V.}~\bibnamefont{Venturi}},
  \bibinfo{author}{\bibfnamefont{J.~I.} \bibnamefont{Cirac}},
  \bibinfo{author}{\bibfnamefont{C.~J.} \bibnamefont{Williams}},
  \bibnamefont{and} \bibinfo{author}{\bibfnamefont{P.}~\bibnamefont{Zoller}},
  \bibinfo{journal}{PRL} \textbf{\bibinfo{volume}{89}}, \bibinfo{pages}{040402}
  (\bibinfo{year}{2002}).

\bibitem[{\citenamefont{Jaksch et~al.}(1999)\citenamefont{Jaksch, {H.-J.
  Briegel}, {J. I. Cirac}, Gardiner, and Zoller}}]{Jaksch:entanglementlattices}
\bibinfo{author}{\bibfnamefont{D.}~\bibnamefont{Jaksch}},
  \bibinfo{author}{\bibnamefont{{H.-J. Briegel}}},
  \bibinfo{author}{\bibnamefont{{J. I. Cirac}}},
  \bibinfo{author}{\bibfnamefont{C.~W.} \bibnamefont{Gardiner}},
  \bibnamefont{and} \bibinfo{author}{\bibfnamefont{P.}~\bibnamefont{Zoller}},
  \bibinfo{journal}{Phys. Rev. Lett.} \textbf{\bibinfo{volume}{82}},
  \bibinfo{pages}{1975} (\bibinfo{year}{1999}).

\bibitem[{\citenamefont{Mandel et~al.}(2003)\citenamefont{Mandel, Greiner,
  Widera, Rom, {T. W. H\"ansch}, and Bloch}}]{Mandel:entanglementlattices}
\bibinfo{author}{\bibfnamefont{O.}~\bibnamefont{Mandel}},
  \bibinfo{author}{\bibfnamefont{M.}~\bibnamefont{Greiner}},
  \bibinfo{author}{\bibfnamefont{A.}~\bibnamefont{Widera}},
  \bibinfo{author}{\bibfnamefont{T.}~\bibnamefont{Rom}},
  \bibinfo{author}{\bibnamefont{{T. W. H\"ansch}}}, \bibnamefont{and}
  \bibinfo{author}{\bibfnamefont{I.}~\bibnamefont{Bloch}},
  \bibinfo{journal}{Nature (London)} \textbf{\bibinfo{volume}{425}},
  \bibinfo{pages}{937} (\bibinfo{year}{2003}).

\bibitem[{\citenamefont{Takamoto et~al.}(2005)\citenamefont{Takamoto, {F.-L.
  Hong}, Higashi, and Katori}}]{Takamoto:LatticeClock}
\bibinfo{author}{\bibfnamefont{M.}~\bibnamefont{Takamoto}},
  \bibinfo{author}{\bibnamefont{{F.-L. Hong}}},
  \bibinfo{author}{\bibfnamefont{R.}~\bibnamefont{Higashi}}, \bibnamefont{and}
  \bibinfo{author}{\bibfnamefont{H.}~\bibnamefont{Katori}},
  \bibinfo{journal}{Nature (London)} \textbf{\bibinfo{volume}{435}},
  \bibinfo{pages}{321} (\bibinfo{year}{2005}).

\bibitem[{\citenamefont{Gerbier et~al.}(2006)\citenamefont{Gerbier, {S.
  F\"olling}, Widera, Mandel, and Bloch}}]{Gerbier:NumberSqueezingLattice}
\bibinfo{author}{\bibfnamefont{F.}~\bibnamefont{Gerbier}},
  \bibinfo{author}{\bibnamefont{{S. F\"olling}}},
  \bibinfo{author}{\bibfnamefont{A.}~\bibnamefont{Widera}},
  \bibinfo{author}{\bibfnamefont{O.}~\bibnamefont{Mandel}}, \bibnamefont{and}
  \bibinfo{author}{\bibfnamefont{I.}~\bibnamefont{Bloch}},
  \bibinfo{journal}{Phys. Rev. Lett.} \textbf{\bibinfo{volume}{96}},
  \bibinfo{pages}{090401} (\bibinfo{year}{2006}).

\bibitem[{\citenamefont{Roberts and Burnett}(2003)}]{Roberts:ProbingMott}
\bibinfo{author}{\bibfnamefont{D.~C.} \bibnamefont{Roberts}} \bibnamefont{and}
  \bibinfo{author}{\bibfnamefont{K.}~\bibnamefont{Burnett}},
  \bibinfo{journal}{Phys. Rev. Lett.} \textbf{\bibinfo{volume}{90}},
  \bibinfo{pages}{150401} (\bibinfo{year}{2003}).

\bibitem[{\citenamefont{Ryu et~al.}(2005)\citenamefont{Ryu, Du, Yesilada,
  Dudarev, Wan, Niu, and Heinzen}}]{Ryu:photoassociationlattice}
\bibinfo{author}{\bibfnamefont{C.}~\bibnamefont{Ryu}},
  \bibinfo{author}{\bibfnamefont{X.}~\bibnamefont{Du}},
  \bibinfo{author}{\bibfnamefont{E.}~\bibnamefont{Yesilada}},
  \bibinfo{author}{\bibfnamefont{A.~M.} \bibnamefont{Dudarev}},
  \bibinfo{author}{\bibfnamefont{S.}~\bibnamefont{Wan}},
  \bibinfo{author}{\bibfnamefont{Q.}~\bibnamefont{Niu}}, \bibnamefont{and}
  \bibinfo{author}{\bibfnamefont{D.~J.} \bibnamefont{Heinzen}}
  (\bibinfo{year}{2005}), \eprint{cond-mat/050801}.

\bibitem[{\citenamefont{Klinner et~al.}(2006)\citenamefont{Klinner, Lindholdt,
  Nagorny, and Hemmerich}}]{Klinner:NormalModeSplitting}
\bibinfo{author}{\bibfnamefont{J.}~\bibnamefont{Klinner}},
  \bibinfo{author}{\bibfnamefont{M.}~\bibnamefont{Lindholdt}},
  \bibinfo{author}{\bibfnamefont{B.}~\bibnamefont{Nagorny}}, \bibnamefont{and}
  \bibinfo{author}{\bibfnamefont{A.}~\bibnamefont{Hemmerich}},
  \bibinfo{journal}{Phys. Rev. Lett.} \textbf{\bibinfo{volume}{96}},
  \bibinfo{pages}{023002} (\bibinfo{year}{2006}).

\bibitem[{\citenamefont{Buck and Sukamar}(1981)}]{twophotonJC1}
\bibinfo{author}{\bibfnamefont{B.}~\bibnamefont{Buck}} \bibnamefont{and}
  \bibinfo{author}{\bibfnamefont{C.~V.} \bibnamefont{Sukamar}},
  \bibinfo{journal}{Phys. Lett.} \textbf{\bibinfo{volume}{81A}},
  \bibinfo{pages}{132} (\bibinfo{year}{1981}).

\bibitem[{\citenamefont{Sukamar and Buck}(1981)}]{twophotonJC2}
\bibinfo{author}{\bibfnamefont{C.~V.} \bibnamefont{Sukamar}} \bibnamefont{and}
  \bibinfo{author}{\bibfnamefont{B.}~\bibnamefont{Buck}},
  \bibinfo{journal}{Phys. Lett.} \textbf{\bibinfo{volume}{83A}},
  \bibinfo{pages}{211} (\bibinfo{year}{1981}).

\bibitem[{\citenamefont{Mekhov et~al.}(2006)\citenamefont{Mekhov, Maschler, and
  Ritsch}}]{Mekhov:ProbingLattice}
\bibinfo{author}{\bibfnamefont{I.~B.} \bibnamefont{Mekhov}},
  \bibinfo{author}{\bibfnamefont{C.}~\bibnamefont{Maschler}}, \bibnamefont{and}
  \bibinfo{author}{\bibfnamefont{H.}~\bibnamefont{Ritsch}}
  (\bibinfo{year}{2006}), \eprint{quant-ph/0610073v1}.

\end{thebibliography}
\end{document}